\documentclass{JINST}

\title{Environmental Effects on TPB Wavelength-Shifting Coatings}

\author{C.S. Chiu$^a$, C. Ignarra\thanks{Corresponding
author.}~, L. Bugel, H. Chen, J.M. Conrad, B.J.P. Jones, T. Katori, I. Moult\\
\llap{$^a$}Massachusetts Institute of Technology,\\
  77 Massachusetts Avenue, Cambridge MA 02139, USA\\
  E-mail: \email{cschiu@mit.edu}}

\abstract{The scintillation detection systems of liquid argon time projection chambers (LArTPCs) require wavelength shifters to detect the 128 nm scintillation light produced in liquid argon. Tetraphenyl butadiene (TPB) is a fluorescent material that can shift this light to a wavelength of 425 nm, lending itself well to use in these detectors. We can coat the glass of photomultiplier tubes (PMTs) with TPB or place TPB-coated plates in front of the PMTs. 
  In this paper, we investigate the degradation of a chemical TPB coating in a laboratory or factory environment to assess the viability of long-term TPB film storage prior to its initial installation in an LArTPC. We present evidence for severe degradation due to common fluorescent lights and ambient sunlight in laboratories, with potential losses at the 40\% level in the first day and eventual losses at the 80\% level after a month of exposure. We determine the degradation is due to wavelengths in the UV spectrum, and we demonstrate mitigating methods for retrofitting lab and factory environments.
}

\keywords{Tetraphenyl Butadiene, TPB, MicroBooNE, Liquid Argon Time Projection Chamber, LArTPC}

\begin{document}

\section{Introduction}

Liquid argon time projection chambers (LArTPCs) are appealing due to their excellent energy resolution and particle reconstruction capabilities. In these detectors, we apply a voltage across a volume of liquid argon (LAr). A charged particle traveling through the argon will ionize the atoms along its path, producing a track of ionization electrons that then drift towards the wire planes of the TPC due to the applied electric field. A typical design \cite{MicroBooNE, icarus, ArgoNeuT}  has three wire planes for detection: two induction planes, which measure the current induced in the wires by the approach of the ionization electrons, and one collection plane. The $dE/dx$ of the charged particles can be reconstructed from this information, as can their 3D tracks.  This allows for excellent particle identification of tracks and showers, with millimeter-level spatial resolution \cite{MicroBooNE}.

This technology is relatively new and its development is rapidly progressing. Previous LArTPCs and successful training grounds for technology in particle physics include ICARUS \cite{icarus} and ArgoNeuT \cite{ArgoNeuT}. Upcoming experiments include MicroBooNE, which is scheduled to begin taking data in 2014. These are all precursors to the ultra-large detectors proposed for future long-baseline experiments, such as the Long-Baseline Neutrino Experiment (LBNE) \cite{LBNE}, and will help us to determine the viability of LArTPCs as an option for this large-scale neutrino program.

Another feature of this technology is that charged particles passing through LAr produce scintillation light which can be detected to provide useful additional information. It is advantageous for the experiment to use this light in trigger and veto systems: a TPC takes order 1 ms to detect an ionization electron, compared to order 6 ns to detect scintillation light. Also the light can provide the interaction time $T_0$ for track reconstruction. This is especially important for non-beam physics because, unlike with beam events, we do not have the spill timing to obtain a sufficiently accurate $T_0$ \cite{MicroBooNETDR}. Unfortunately, LAr scintillation light is produced at 128 nm and, therefore, must be shifted to visible wavelengths for detection. This is typically performed using the wavelength shifter TPB.

As detector size grows, construction and installation will inevitably require larger spaces and production-based environments. This paper explores how aspects of a typical detector assembly environment can affect the TPB used in LArTPC light collection systems and suggests practical methods to mitigate these effects. We begin by describing a robust coating method for applying TPB that can survive handling in a large assembly area without imposing excessive costs. Next we focus on the effects of exposure to ambient laboratory light on this TPB coating. We then turn to other light spectra and explore their effects on TPB degradation. We end with a set of practical recommendations for constructing and handling light collection systems for LArTPCs.

\section{Light Detection in Liquid Argon}

Scintillation light can be produced via a fast or slow path \cite{LArScint, LarScint2}. The fast path has a time constant of 6 ns and produces about 25\% of the total scintillation light, whereas the slow path has a time constant of 1600 ns and produces the remaining 75\% of light. They both occur when a charged particle passing through the detector results in the formation of argon excimers, each of which then decays into two individual argon atoms and releases a 128 nm photon. The main difference lies in the fact that the charged particle initially excites the argon in the fast path, and ionizes it in the slow path. Additionally, the excimer of fast scintillation is in the singlet state before decaying into the two argon atoms and one photon; whereas slow scintillation excimers begin in the triplet state before decaying into the singlet state, which then produces the two argon atoms and a 128 nm photon.

The scintillation light, at 128 nm, lies in the vacuum UV range and thus cannot pass through many materials, including air and glass. Its wavelength must be shifted to allow for detection. TPB lends itself well to this task given that it absorbs UV light and re-emits it in the visible spectrum at a peak wavelength of 425 nm. We can implement light conversion through applying a TPB film to the glass of the PMTs \cite{icarus}, or to separate plates \cite{MicroBooNETDR} or lightguides \cite{lightguidepaper} which can then be placed in front of the PMTs. This allows any light reaching the PMTs to be shifted to a wavelength that can pass through the PMTs' glass envelopes.

\section{TPB-coated Plates and Test Stand Used in This Study} \label{sec:coatings}

\subsection{Application of TPB to Acrylic Plates}

The efficiency, in terms of photons out per photon in, for re-emission of light absorbed at 128 nm has been determined to be 120\% for a film of pure TPB made through vacuum evaporation \cite{tpbeff}. This follows energy conservation because the re-emitted photons have a lower average energy than the incident light. However, we have observed that evaporative coating on acrylic handled in a lab environment is not resilient \cite{MicroBooNETDR}, presumably due to absorption of water by the TPB \cite{HahnPrivateComm}. Construction of a large light collection system would be extremely expensive if the acrylic must be maintained free of water. Therefore, we have looked to other, more resilient coatings.  

We are analyzing a number of different coating mixtures, and they will be reported elsewhere. For this study, we used a coating of 50\% TPB and 50\% polystyrene (PS) by mass, which will be used in the MicroBooNE experiment. This has been measured to have about 50\% of the efficiency of the evaporative coating \cite{MicroBooNETDR}. We found that using mixtures such as these makes the plates more durable and will be much more cost-effective for large systems.

\begin{figure}[tb]
\centering
\includegraphics[width=0.4\textwidth]{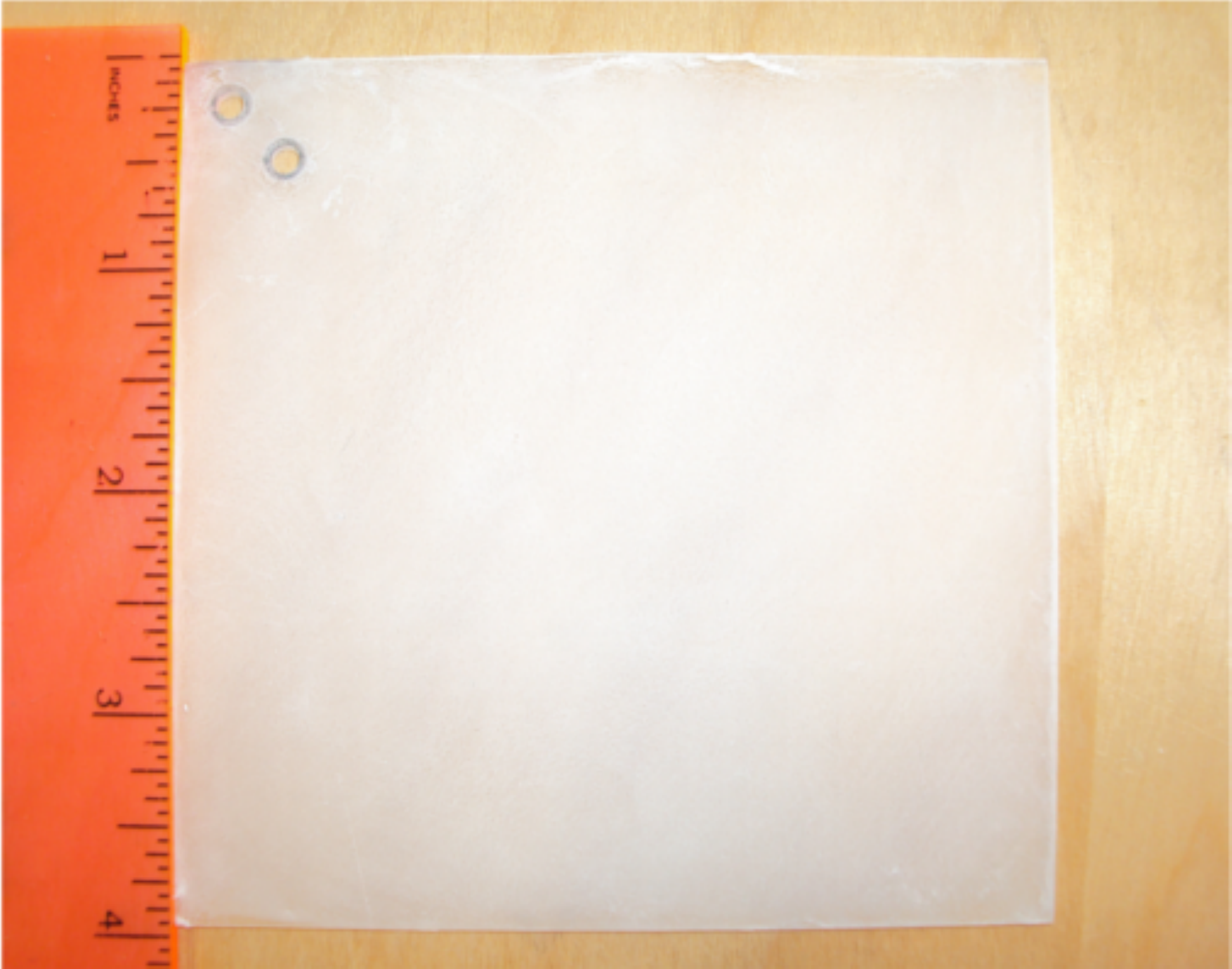}
\caption{A typical plate used for our studies. The acrylic plate is clear, and the TPB coating has a milky appearance. Two holes were drilled into the plate for attachment to our testing apparatus (Figure \protect \ref{fig:apparatus}).}  \label{fig:plate}
\end{figure}

To adhere the TPB to each plate, we created a TPB solution consisting of a 1:1:43.3 ratio by mass of PS to TPB to toluene. Each acrylic plate required three coatings of the TPB solution (Figure \ref{fig:plate}). For this study, each plate measured approximately 4 inches square; however, larger plates are planned for MicroBooNE. For each coating, we applied about 2.5 to 3 mL of solution onto the test plate and distributed it evenly with a brush by hand. The solution temporarily dissolved any previous coatings from the plate, leaving a more concentrated solution of TPB+PS in toluene. The toluene then evaporated, leaving behind a thin film of TPB embedded in PS. Because they were applied by hand, there were inevitable inhomogeneities in the coatings. Consequently, we tested the performance of two locations on each plate when monitoring the possible degradation of TPB in a laboratory environment.

\subsection{Testing Plate Performance} \label{sec:teststand}

\begin{figure}[tb]
  \centering
  \includegraphics[width=0.75\textwidth]{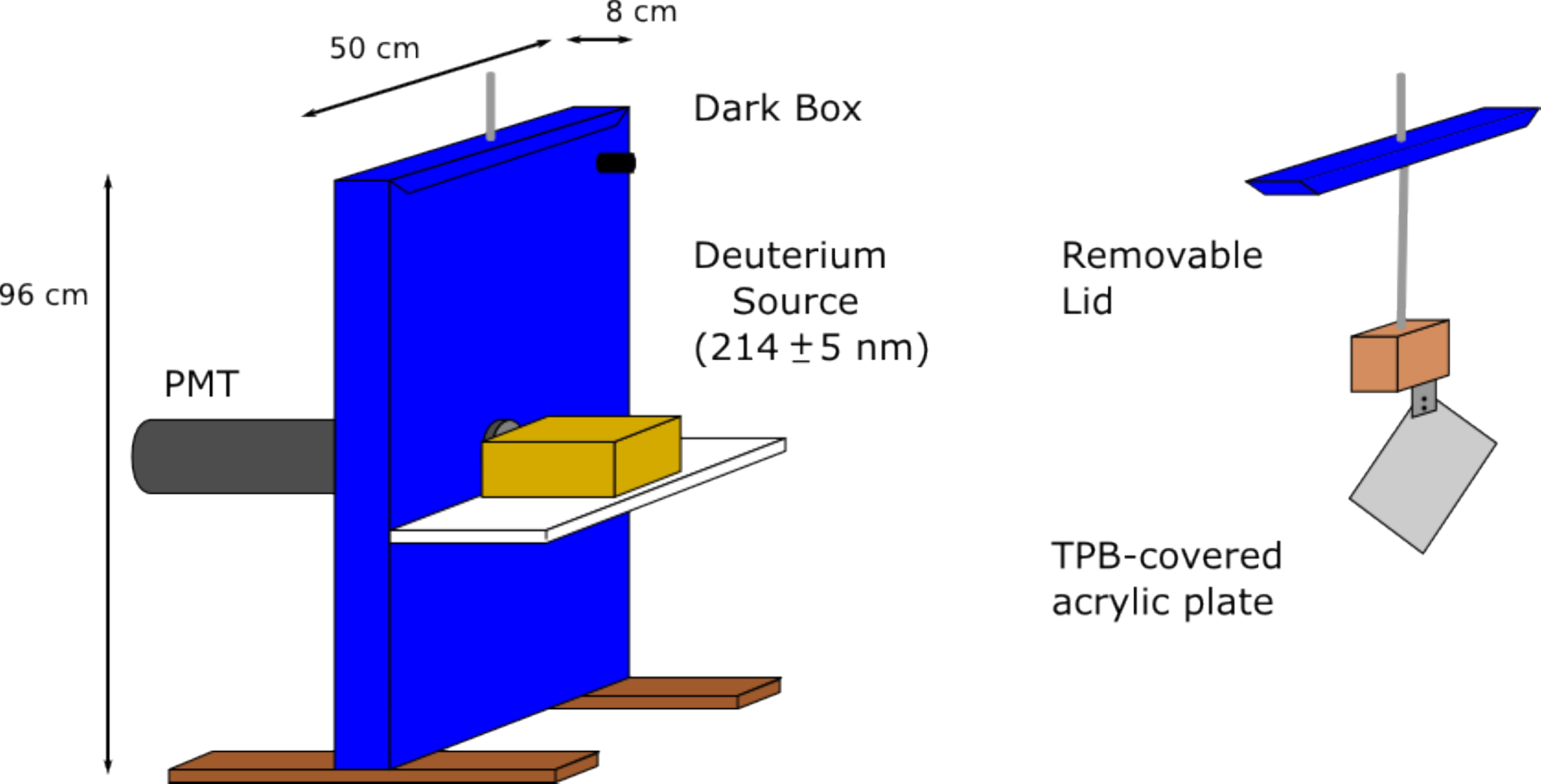}
  \caption{The plate performance test stand consisted of a dark box with a PMT and deuterium source. The plate could be lifted independently of the lid to test different locations on the plate.} \label{fig:apparatus}
\end{figure}

The following tests used the apparatus shown in Figure \ref{fig:apparatus} to measure plate response. It comprised a 50 cm x 8 cm x 96 cm dark box with a PMT attached to one side. On the opposite side we placed a StellarNet Inc. SL3 deuterium lamp filtered to emit wavelengths at $214 \pm 5$ nm. The lid was fitted with a rod which extended into the box, and to which we could attach an acrylic plate. To test different locations on the plate, it could be lifted independently from the lid to a consistent height. Plate performance was quantified using the number of PMT pulses obtained above a threshold of 30 mV in a designated time interval. Typical measurements ranged from about 12,000 counts/sec for a degraded plate to 70,000 for a newly made plate. The dark rate for pulses over 30 mV equaled $260. \pm 5$ counts, and thus could be regarded as negligible.

\section{Basic Lighting Conditions} \label{sec:basicLight}

\subsection{Setup} \label{subsec:basicProc}

We began by examining TPB film performance after extended exposure to four common laboratory lighting conditions. After applying the TPB film and recording their initial performance levels, we placed groups of three plates in one of four conditions. Three of these groups were placed into lab humidity, but each was exposed to a different kind of light: common fluorescent light, amber light, or no light. The fourth condition kept plates in the dark, but also contained reduced humidity levels to examine whether humidity was a concern for plate degradation in the long run.

The plates were kept in their respective environments for about a month, and their performances were initially monitored every one or two days. TPB plate performance was assessed by averaging the PMT count rate from the testing apparatus with the plate in place, sampled for two different locations on the plate. However, we have noticed downward drifts in PMT sensitivity by about 8\% in five minutes, as well as in the deuterium source output over several hours. Therefore, we let the PMT stabilize for at least five minutes and the source for at least three hours before use. We accounted for any remaining fluctuations by introducing a reference plate: a TPB-coated plate that has already reached an approximately steady state degradation and, therefore, would not change in performance over the course of our study.  This reference plate was also kept in the dark to ensure its stability. By measuring the performance of the reference plate before and after each testing session, we could normalize each plate relative to the reference and track this normalized performance over time.

\subsection{Results}

\begin{figure}[tb]
  \centering
 \includegraphics[width=0.75\textwidth]{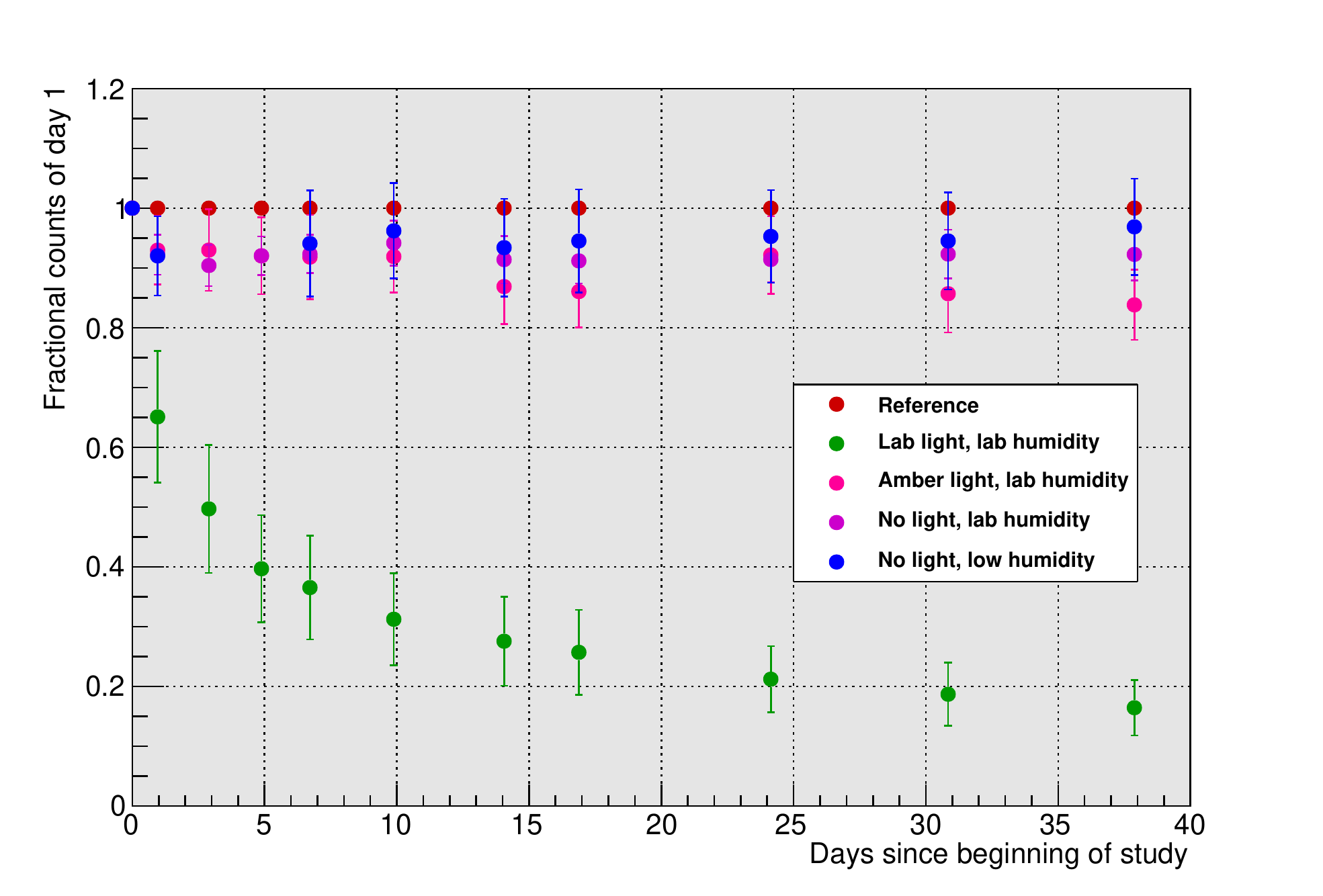} 
  \caption{Degradation of TPB films due to various lighting conditions, as fractions of their performance on the day that they were created. Data has been normalized to a reference plate.}
  \label{fig:lightDegrad}
\end{figure}

The average count rates per plate and per storage condition were computed for each day and normalized to the reference plate averages. Plate performance, expressed as a percentage of Day 1 performance, is plotted for each plate condition in Figure \ref{fig:lightDegrad}. Sources of error included fluctuations in the testing apparatus and plate inhomogeneity.

Our results are quite striking; within one day, we observed roughly 35\% degradation due to common laboratory lighting conditions. In comparison, plates exposed to amber light or no light experienced less than 10\% degradation.  We believe the initial drop in the dark- and amber-stored plates was due to the light exposure from creation to initial testing. Furthermore, after five weeks these samples still exhibited similar performance levels, suggesting that degradation over these timescales due to amber light, no light, or lab humidity is not of major concern. Five weeks of exposure to ambient lab light, however, was quite detrimental: our plates demonstrated less than 20\% of their original wavelength-shifting ability. Thus we observe the harm done to our TPB films due to unfiltered light, with detrimental effects on a timescale of less than a day.

\section{Exposure to Restricted Light Spectra}

\subsection{Setup}

\begin{figure}[tb]
  \centering
  \includegraphics[width=0.75\textwidth]{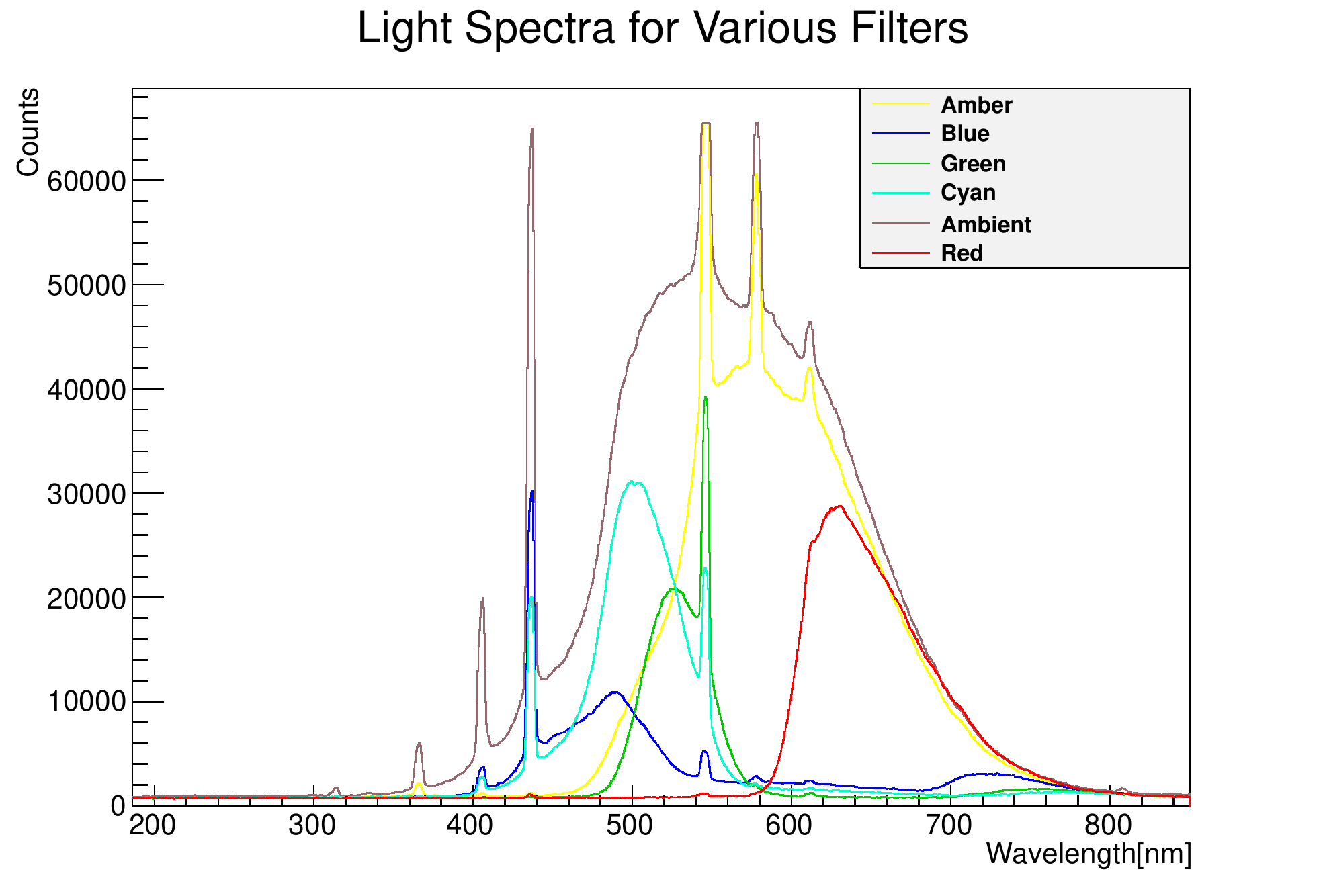} 
  \caption{Spectra of full spectrum fluorescent light after it passes through one of the colored filters. The peak at about 440 nm is due to a spectrometer artifact; however, all the others arise from emission lines in the lamp.}
  \label{fig:colorSpect}
\end{figure}

We introduced five ranges of light wavelengths to test TPB-coated plates. Each condition exposed the plates to a different spectrum of light by means of a full spectrum fluorescent light shone through various light filters. Each filter was centered around either amber, blue, green, cyan, or red, although each filter also let through some light at other wavelengths. All of the spectra under consideration are displayed in Figure \ref{fig:colorSpect}.

We tested the plates every three to four days for approximately a month using the method described in section \ref{subsec:basicProc}. After sixteen days, we increased the intensity of light to which all plates were exposed. This was done to better distinguish whether TPB films were degrading slowly or not at all. We did this by halving the distance between plates and the fluorescent light source.

\subsection{Results}

\begin{figure}[tb]
  \centering
  \includegraphics[width=0.75\textwidth]{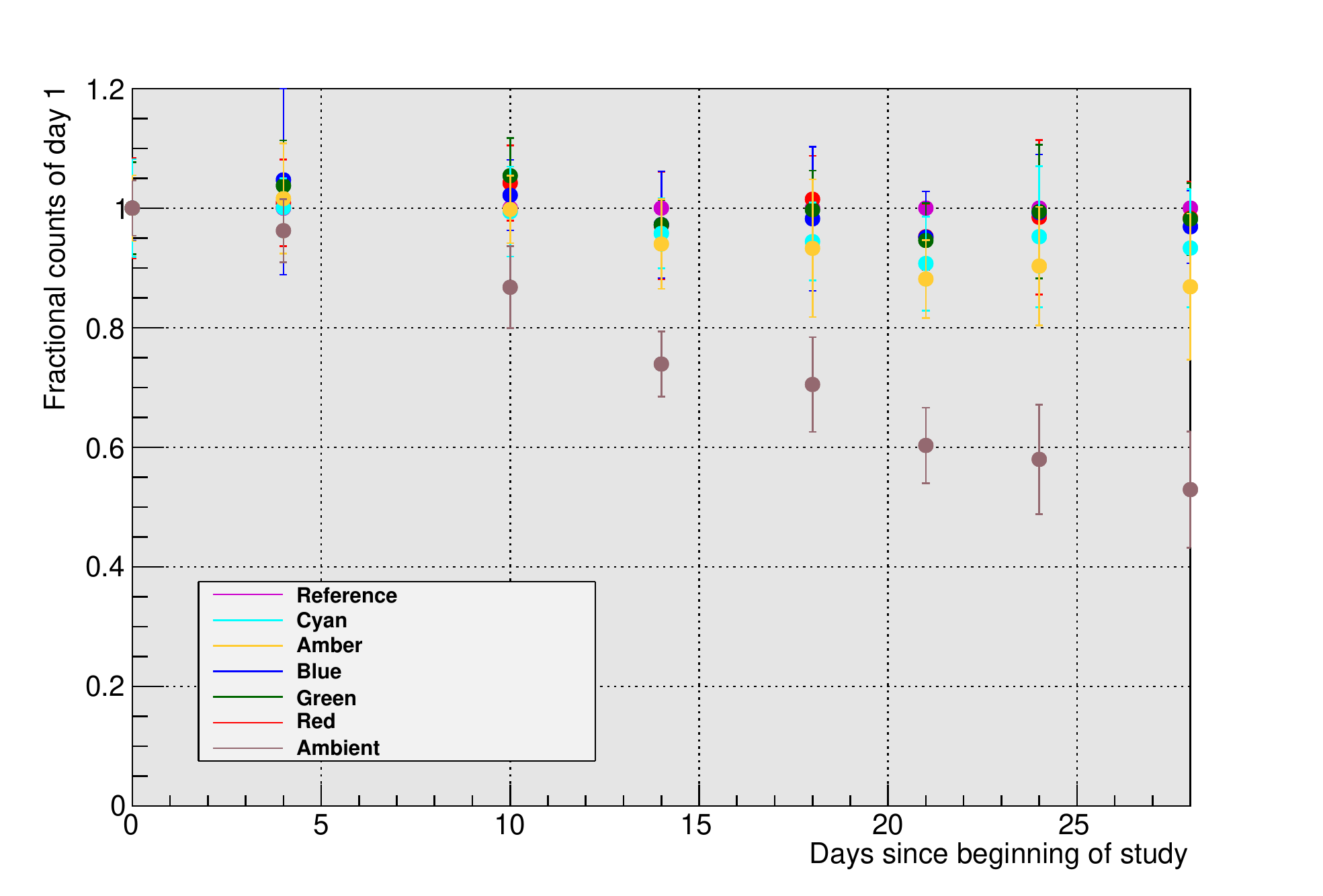} 
  \caption{Fractional initial performance of TPB-coated plates placed under each type of filtered light, monitored over one month.}
  \label{fig:colorDegrad}
\end{figure}

Figure \ref{fig:colorDegrad} plots the percent degradation over this time period and suggests that wavelengths of light in the visible spectrum do little damage to TPB films. Indeed, plates exposed to light within the visible spectrum seem to have degraded by 15\% at most over four weeks, while the full spectrum fluorescent light reduced TPB film performance by about 50\% over that time. This result, combined with our conclusions from Section \ref{sec:basicLight}, seems to suggest that light in the ultraviolet region plays the largest role in TPB degradation. In particular, wavelengths longer than 700 nm are less likely to do harm than wavelengths shorter than 400 nm, due to their lower energies. This hypothesis was confirmed by testing plates with UV-blocking filters and observing a decrease in degradation similar to that seen when using any of the colored filters. Visible light may have a small detrimental effect on TPB, because increasing the intensity of light increased degradation. Another possibility is that UV light may have leaked through the low-wavelength cutoff of the colored filters.

\section{UV-blocking Materials as a Preventative Measure}

\subsection{Setup}

\begin{figure}[tb]
  \centering
  \includegraphics[width=0.75\textwidth]{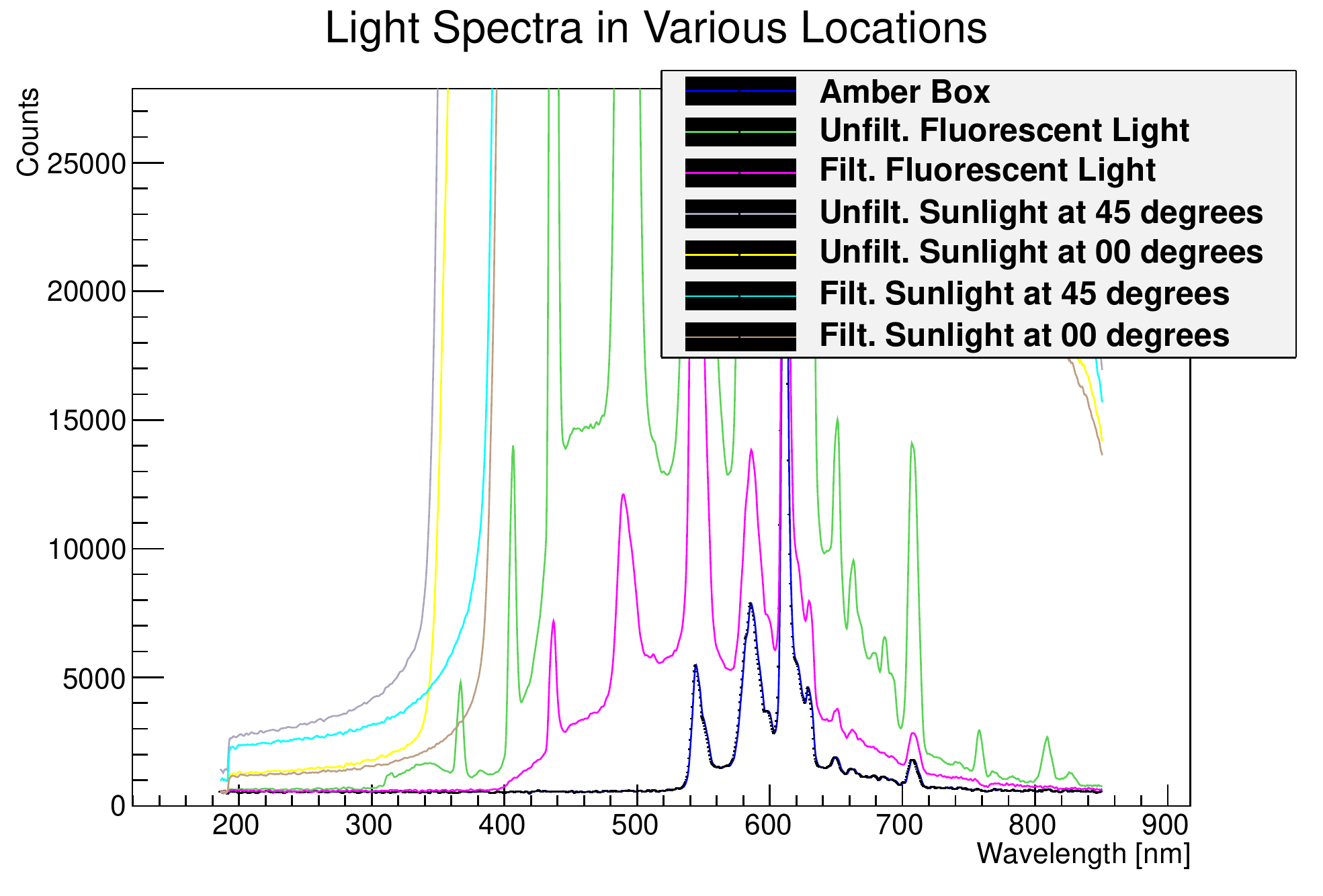} 
  \caption{Light spectra of the light sources with and without UV blockers in place. Notice also that the viewing angle has a large effect on each spectrum. Because distance from the light source largely determines light intensity, the number of counts does not provide any useful information. Rather, we are simply interested in the cutoff wavelengths between 300 and 450 nm.}
  \label{fig:UVblockSpect}
\end{figure}

Inferring that wavelengths in the UV spectrum are most likely the main cause of TPB film degradation, we looked to UV-blocking materials to mitigate the harm done. The two possible sources of UV light in our laboratory are the UV components of common fluorescent light and of ambient sunlight. We targeted each source separately: McMaster-Carr sells UV-blocking sheaths that fit over fluorescent lights, and we have found commercially available UV-blocking films for installation on our windows. The spectrum of each light source with and without each UV-blocker is shown in Figure \ref{fig:UVblockSpect}. We then examined the ability of each to reduce or prevent TPB degradation.

These preventative measures were evaluated separately for the fluorescent lights and sunlight. We monitored TPB degradation due to filtered and unfiltered sunlight without fluorescent lights, where plates were placed on the windowsill in direct sunlight. Similarly, we examined degradation due to filtered and unfiltered fluorescent lights with no ambient sunlight. Measurements to quantify performance were taken with the test stand described in Section \ref{sec:teststand}.

\subsection{Results}

For TPB-coated plates placed on a windowsill under direct sunlight and no fluorescent light, we noticed less degradation when the UV-blocking films were installed on the windows. After about 3.5 hours of exposure to these two conditions, we report about 15\% degradation in filtered sunlight and 47\% degradation in unfiltered sunlight. Thus, it seems that although the UV-blocking films did not completely stop TPB degradation, they were very helpful in ameliorating the situation. We suspect that the magnitude of degradation from sunlight also depended on environmental factors such as time of day, season, and cloud cover. We note that, in general, plates should be stored out of direct sunlight, with or without UV filters. 

A similar effect occurred for plates placed under fluorescent lights with and without the UV-blocking sheaths. After approximately 75.4 hours, plates under unfiltered fluorescents degraded by about 15\%, whereas plates under filtered fluorescents degraded by about 4\%. As with sunlight, we believe that proximity to the fluorescent light source influenced degradation as well. Indeed, we observed this to be the case in the colored filters study. From these results, it seems that the main culprit in our study from Section \ref{sec:basicLight} was the sunlight.  We cannot draw any rigorous conclusions regarding the comparison between sunlight and fluorescent light because there are many unmeasured factors, such plate location and angle with respect to each light source.

These results should be used only to compare degradation with and without each of the UV-blockers, and as general indicators of degradation. It is difficult to quantify exactly the amount or wavelength distribution of light reaching a TPB-coated plate placed in any given location in the lab. Not only can sunlight change unpredictably, but proximity and angle to light sources can drastically affect measurements between successive repetitions of this study.

\section{Conclusion}

We have demonstrated the harmful effect of certain wavelengths of light on the degradation of TPB and isolated the most damaging wavelength range to the UV spectrum. On the other hand, we have seen that humidity does not play a significant role in long-run plate degradation. Furthermore, we have determined that UV-blocking film ameliorates the damage from these wavelengths of light. Each of the two types of UV-blocker may reduce degradation by over two-thirds. However, because they cannot completely shield against degradation, we suggest that several other measures be taken to preserve TPB-coated acrylic plates. First, they should always be kept out of direct sunlight. Second, although plates can be exposed to sunlight or fluorescent lights, this should only occur for about an hour at most to minimize any degradation. UV-blocking filters should be installed on windows and lights if possible. Finally, we recommend that although TPB films can be applied in a lighted environment, prepared plates should be stored in an amber box or in the dark until installation. With these precautionary measures, TPB can be used successfully as a wavelength shifter in LArTPCs.

\acknowledgments
The authors thank the National Science Foundation (grant number NSF-PHY-084784) and the Paul E. Gray Fund for the Undergraduate Research Opportunities Program at MIT.

\end{document}